\documentclass[a4paper, 12pt]{article}
\usepackage{epsfig}
\usepackage{amsmath}
\usepackage{amssymb}
\usepackage[authoryear]{natbib}
\usepackage{parskip}
\usepackage[amsmath,thmmarks]{ntheorem}
\usepackage{rotating}
\usepackage{url}

\theorembodyfont{\slshape}

\usepackage{setspace}
\usepackage{algorithm,algorithmic}

\newlength{\tcwname}
\newlength{\tcwnum}
\newcommand{\N}{\mathcal{N}}

\newcommand{\simiid}{\stackrel{\text{iid}}{\sim}}

\newcommand{\E}{\mathbb{E}}

\begin{document}

\title{On Estimating the Ability of NBA Players}
\author{Paul Fearnhead and Benjamin M. Taylor}
\maketitle

\begin{abstract}
This paper introduces a new model and methodology for estimating the ability of NBA players. The main idea is to directly measure how good a player is by comparing how their team performs when they are on the court as opposed to when they are off it. This is achieved in a such a way as to control for the changing abilities of the other players on court at different times during a match. The new method uses multiple seasons' data in a structured way to estimate player ability in an isolated season, measuring separately defensive and offensive merit as well as combining these to give an overall rating. The use of game statistics in predicting player ability will be considered. Results using data from the 2008/9 season suggest that LeBron James, who won the NBA MVP award, was the best overall player. The best defensive player was Lamar Odom and the best rookie was Russell Westbrook, neither of whom won an NBA award that season. The results further indicate that whilst the frequently--reported game statistics provide some information on offensive ability, they do not perform well in the prediction of defensive ability. 
\end{abstract}


\section{Introduction}

The most basic rating systems for professional basketball players are simple (or not so simple) functions of `positive' statistics such as free throw percentage and the number of steals as well as `negative' statistics like the number of turnovers and personal fouls. One example is the computer ranking procedure used to assign the 1998 IBM player award (the function used is detailed in equation 1 of \cite{berri1999}). Although such rating systems yield some information on player ability, they usually offer neither a justification for the functional form, for the choice of player statistics used, for the `value' ascribed to each statistic in calculating the rating nor an estimate of the precision.

One alternative is to model an outcome variable (like whether or not the team won) as a function of game or player statistics, for example by using a regression model. Provided the data is well described according to the chosen model, standard statistical techniques may be applied to overcome and provide answers to the issues mentioned above. One of the earliest published attempts to model professional basketball team performance was by \cite{zak1979} in which a Cobb--Douglas production function was used to model the ratio of final scores against game level statistics including those four mentioned above. Although this paper did not seek to rate individual players, the authors mention that such an approach is feasible using their method.

\cite{berri1999} models team wins as a linear combination of player and game level statistics, the idea being to learn which statistics are valuable in predicting team wins. Unfortunately the problem with ascribing team wins to individual performances over entire games ignores some important information, namely that at any one time there are only 10 active players. Although this issue is addressed in his paper by adjusting for the time spent by each player on court, there is still some loss of information as only certain combinations of players meet each other during the game. Moreover, Berri's method though ingenious, is somewhat intricate and difficult to justify at a methodological level: it is still not clear why the range of statistics considered should be preferred over another candidate set. The reliance on these statistics means that methods such as Berri's tend to be overly complicated; it is the opinion of the present authors that it does not matter \emph{how} a net point difference is achieved, simply that it \emph{is} achieved.

Thanks to modern game charting techniques and records, much more information is now available to researchers \citep{kubatko2007}. Not only are player--level summary statistics available, but some organisations provide time--lines of important events during each game. With this information, it is possible to infer which players were on pitch at any time in the game, thus allowing an alternative modelling perspective. A basketball game consists of a sequence of time intervals in which no substitutions occur. Each of these intervals can be thought of as a `small game' in which the players on court remain constant; the outcome of the game as a whole being the sum of outcomes of the small games. By comparing how the results of these small games depends on the players on the court, the relative ability of the different players can be measured. For example if player A is substituted by player B then the relative ability of these two players is measured by looking at how well their team performs when player A is on the court as compared to player B. 

Formally, the new model introduced in this article measures each player's offensive and defensive ability. The expected number of points per possession for a given team during a `small game' is modelled as a linear function of the offensive ability of the 5 players of that team who are on the court and in possession and the defensive ability of the 5 players on the opposing team, adjusting for home advantage. This approach is similar to that of \cite{rosenbaum2004} and \cite{ilardi2008}. The main difference compared with \cite{rosenbaum2004} is that the author estimates only a combined ability for each player. The model presented here further uses a structured approach to combining information from multiple seasons.

In the model of \cite{ilardi2008} (which has been used by \cite{macdonald2010} in the context of estimating the abilities of NHL players) the home advantage parameter only acts when the home players are in possession, whereas in the present article it also modelled as an effect when they are defending. This article further extends the results of \cite{ilardi2008} by providing a combined measure of player ability and a method for analysing results from a single season informed by data from previous seasons. Furthermore the utility of game statistics in predicting offensive and defensive ability is also considered here.

A summary of the paper is as follows. In section \ref{sect:methods}, the data collection process will be reviewed and the new models for within and between year analyses presented. In section \ref{sect:results} the ability of players at the end of the 2008--2009 season are estimated. In section \ref{sect:playerstats}, the use of game statistics in inferring player ability is addressed. The article ends with a discussion.

\section{Methods\label{sect:methods}}

\subsection{Data Collection}

The information required to fit the proposed model is available from the ESPN website in the form of individual game `play--by--play' and `box score' records \citep{ESPNNBA}. For each regular season match, the box score pages provide summary statistics by player and the play--by--play pages give a detailed record of events over the course of the match. The play--by--play pages consist of a list of important events on court together with the time at which the event occurred. Using the available information on substitutions, together with the list of `starters' (players on court at the start of the match) from the box score page, it is possible to infer exactly which 10 players are on court at any time during the match and furthermore exactly which team is in possession at any time. The latter requires the mild assumption that between time records, possession remains in one team and that all changes of possession are recorded. 

Any quarter or overtime period of a game may therefore be split up into small intervals of time in which the players on court remain constant (this will be taken as the definition of the word `interval' in this paper). The duration of these intervals, the number of possessions for each team and the number of points gained (or conceded) in that space of time can be inferred from the play by play data. In the course of one season, there are of the order of 30000 such intervals, so some data error is to be expected. In the analysis to follow, any interval in which it was not possible to infer the exact 10 players on court was excluded from the analysis; this restriction meant that approximately 10\% of all available data was excluded. Using the game summary information from the box score page, it was possible to cross check inferences from the play--by--play pages for games with complete information (ie those games in which there were no apparent errors), this included total time on pitch by player and total inferred game time; the corroboration between these two sources was excellent.

\subsection{The New Model}

Suppose there are currently $N$ players in the NBA; let $\{\alpha_i\}_{i=1}^N$ and $\{\beta_i\}_{i=1}^N$ denote respectively the attacking and defensive ability of each player. For each game let $H$ be the the home team and $A$ the away team. Suppose the number of intervals in the game is $n$ and for interval $k$, the number of possessions for team $H$ is $n_{Hk}$ and for team $A$ is $n_{Ak}$ (so that $n = \sum_k(n_{Hk}+n_{Ak})$). If team $H$ (resp. $A$) comes into possession in interval $k$, let $y_{Hk}$ (resp. $y_{Ak}$) be the total number of points scored by team $H$ (resp. $A$) in the interval. This gives the model:
\begin{eqnarray}
   100\times{y_{Hk}} &=& n_{Hk} \left(\sum_{i:\text{home player $i$ on court}}\hspace{-3em}\alpha_i\hspace{3em} - \sum_{j:\text{away player $j$ on court}}\hspace{-3em}\beta_j\hspace{2em} + \gamma + \frac\sigma{\sqrt{n_{Hk}}}\epsilon_{Hk} \right), \label{eqn:newmodelH}\\ 
   100\times{y_{Ak}} &=& n_{Ak} \left(\sum_{i:\text{away player $i$ on court}}\hspace{-3em}\alpha_i\hspace{3em} - \sum_{j:\text{home player $j$ on court}}\hspace{-3em}\beta_j\hspace{2em} - \gamma + \frac\sigma{\sqrt{n_{Ak}}}\epsilon_{Ak} \right), \label{eqn:newmodelA}
\end{eqnarray}
where $\epsilon_{Ak}$ and $\epsilon_{Hk}$ are independent standard normal random variables, $\sigma$ is a positive scaling parameter fro the noise and $\gamma$ is a constant representing the home advantage over approximately half of the course of a match (see below).  A typical basketball game has of the order of 100 possessions, hence the function of the multiplicative factor on the LHS of (\ref{eqn:newmodelH}) and (\ref{eqn:newmodelA}) is to scale the estimates of the $\alpha$s and $\beta$s so that they can be interpreted at the more meaningful game level; a similar idea was used by \cite{rosenbaum2004,ilardi2008}.

The interpretation of the parameters is that, for a set of 5 home players $\mathcal K$ and 5 different away players $\mathcal L$, where $\mathcal K,\mathcal L\subset\{1,\ldots,N\}$, assuming $n_{Hk}=n_{Ak}$ for all $k$, the quantity,
\begin{equation*}
   \sum_{k\in\mathcal K} (\alpha_k + \beta_k) - \sum_{l\in\mathcal L} (\alpha_l + \beta_l) + 2\gamma,
\end{equation*}
would be the expected score difference between the home and away players over a period of time approximately equal to that of a typical NBA game without overtime. A possible criticism of the above model is the use of a Gaussian error term, since the chosen outcome variable is always non--negative. The model is justifiable on the basis of a central limit theorem argument: a game consists of a series of around 100 alternating possessions, the outcome of the game being the difference in the sum of points scored by each team in their share of possessions \citep{harville2003}. Although the parameters $\alpha_i$ and $\beta_j$ are not identified by this model (which means that the likelihood of the observed data is unchanged if the same constant is added to each $\alpha_i$ and $\beta_j$) their relative difference \emph{is} estimable, which allows ratings and standard errors to be constructed directly using the posterior density of the unknown parameters.

The parameters $\gamma$ and $\sigma$ were treated as fixed, which is justifiable on the basis that there is a lot of information in the data on these quantities. The other parameters in the model were assumed to be drawn from Gaussian distributions, which represent the variability in offensive and defensive ability of NBA players,
\begin{eqnarray}
   \alpha_i &\sim& \N(\mu_\alpha,\sigma_\alpha^2), \label{eqn:priorspec} \\
   \beta_i &\sim& \N(\mu_\beta,\sigma_\beta^2);\nonumber
\end{eqnarray}
the posterior will therefore also be Gaussian. The fixed parameters and prior hyperparameters were estimated by maximum likelihood (ML) and are given in Table \ref{tab:hyperparameter_estimates}.

\begin{table}[htbp]
   \centering
   \begin{tabular}{c|cccccc}
      Parameter & $\mu_\alpha$ & $\sigma_\alpha$ & $\mu_\beta$ & $\sigma_\beta$ &  $\gamma$ & $\sigma$ \\ 
      Estimate & 9.82 & 2.55 & -9.12 & 1.82 & 1.43 & 106.8
   \end{tabular}
   \caption{Prior hyperparameter estimates. \label{tab:hyperparameter_estimates}}
\end{table}

The interpretation of these parameters is that $\frac{1}{100}(5\mu_\alpha-5\mu_\beta+2\gamma)=0.98$ is the expected number of points scored by a team on their home court (see (\ref{eqn:newmodelH})) in one possession; the standard error is $1.1$. 

One point of interest from Table \ref{tab:hyperparameter_estimates} is that the estimated prior hyperparameter for the variance of the $\beta$s is smaller than that for the $\alpha$s, this indicates that under the proposed model, players are more similar in terms of defensive ability than in terms of offensive ability. The ML estimate of $\gamma$ suggests that there is a home--court advantage of around three points, which was similar to the figure of 3.6 found by \cite{entine2008}, using data from two seasons.

\subsection{Allowing Player Strength to Vary Over Time \label{sect:betweenyear}}

A further extension is to consider the strength of players over a number of years. This is most simply achieved by using the model described above in equations (\ref{eqn:newmodelH}) and (\ref{eqn:newmodelA}) for each year. For the first year the priors given by (\ref{eqn:priorspec}) are used, then for subsequent years, a model is introduced to describe how a player's ability may change from year to year. Let $\alpha^{(t)}_i$ and $\alpha^{(t-1)}_i$ be the respectively the offensive ability of player $i$ at the start of season $t$ and at the end of season $t-1$; define similar terms for the defensive abilities, $\beta_i$. A simple way to allow for a change in ability between seasons is to assume,
\begin{equation}\label{eqn:betweenyearmodel}
   \alpha_{i}^{(t)}= p\alpha^{(t-1)}_i + (1-p)\mu_\alpha + s_\alpha\epsilon_i,\qquad  \beta_{i}^{(t)}= p\beta^{(t-1)}_i + (1-p)\mu_\beta + s_\beta\epsilon_i,
\end{equation}
where $s_\alpha,s_\beta>0$ and $\epsilon_i\simiid\N(0,1)$. If a new player starts in season $t$, then he is assigned a normal prior, as per (\ref{eqn:priorspec}). The two effects of this end--of--year transition are firstly that $p$ shrinks each parameter estimate towards the prior mean, $\mu_\alpha$ or $\mu_\beta$; secondly, by adding either $\N(0,s_\alpha^2)$ or $\N(0,s_\beta^2)$ noise, the uncertainty in each of the estimates is increased whilst some of the correlation structure learned to--date is preserved (the off--diagonal correlations being shrunk towards zero). Increasing the uncertainty in the estimates at the end of the year is important because it allows each player's ability to vary over time: one might expect a player to (at least initially) improve with experience, but also they may have been injured during the previous season or change teams at the end. Using data from the 2006/2007 -- 2008/2009 seasons, the respective maximum likelihood estimates of $p$, $s_\alpha$ and $s_\beta$ were $0.83$, $1.23$ and $0.59$.

\subsection{Estimating Ability in a Single Season\label{sect:reducedvariance}}

It is often of interest to estimate how well each player has performed in just the most recent season, one example being the decision as to who should receive the annual NBA awards. Whilst it is possible to obtain such estimates using multiple years data and the methodology discussed thus far, the estimates of ability obtained from these models can be heavily influenced by performance in previous seasons (consider the example of the combined ability of Kevin Garnett in the results section). For rookie players, the methodology discussed does not present any problems as there is no information from previous years on the ability of these players. For non--rookie players on the other hand, data from earlier seasons is still useful because it enables better estimates of how good the other players are, but the manner in which this information should be included requires care. In this section, a method for handling information from previous seasons is presented, this method may be used to estimate the abilities of non--rookie players in a particular season.

Suppose interest is in estimating how well player $i$ performed in the most recent season. The approach suggested here is to use data from the most recent season to estimate the parameters associated with player $i$; but data from all seasons to estimate the parameters associated with the other players. The simplest way of implementing this is to analyse all seasons' data, as in Section \ref{sect:betweenyear}, but changing the prior distribution for $\alpha_i$ and $\beta_i$ for the most recent season so that it is independent of data in previous seasons. If $t$ denotes the most recent season, this is achieved by imposing the prior $\alpha_i^{(t)} \sim \N(\mu_\alpha,\sigma_\alpha^2)$ and $\beta_i^{(t)} \sim N(\mu_\beta,\sigma_\beta^2)$; rather than using (\ref{eqn:betweenyearmodel}). The posterior distribution for $\alpha_i^{(t)}$ and $\beta_i^{(t)}$ under this model is a measure of how well player $i$ performed in that season. This method reduces the standard errors of parameter estimates compared to those that would be obtained with a single season's data under model (\ref{eqn:newmodelH})--(\ref{eqn:newmodelA}). 

The model in this section will henceforth be referred to as model (\ref{sect:reducedvariance}).

\section{Results\label{sect:results}}

\subsection{2008/2009 Season Results}

In this section the 2009 season results will be presented. Tables \ref{tab:offensiveratings}, \ref{tab:defensiveratings} and \ref{tab:combinedratings} give respectively offensive, defensive and combined player ratings, with standard errors in parentheses, and rankings for the top ten players under the three scenarios described. The ratings and rankings in the second column of these tables correspond to model (\ref{eqn:newmodelH})--(\ref{eqn:newmodelA}) using the 2007--2009 data; in the third column are results from the same model, but only using the 2009 data; and the fourth column gives the results from model (\ref{sect:reducedvariance}). The parameter estimates in these tables are centred.

Under model (\ref{eqn:newmodelH})--(\ref{eqn:newmodelA}), the 2007--2009 data gives greater accuracy in predicting player ability compared to using the 2009 data only; however it may not always be appropriate to use information from previous years. The estimates in the third column are therefore the best estimates of player abilities using only 2009 season information, and those in the fourth are the overall best estimates for this season using all available information. 

For the 2009 season only (columns three and four), there was much similarity between the results from model (\ref{eqn:newmodelH})--(\ref{eqn:newmodelA}) and model (\ref{sect:reducedvariance}); both models identifying the same top three players in the same order for the combined ratings. One of the interesting point from these tables is that using the 2007--2009 data, Kevin Garnett (the most highly paid player in 2009) is identified as the number 3 player, but by using only the 2009 data, he ranks 13th/12th. The likely cause of this is the fact that during a game against the Utah Jazz, Garnett strained his right knee: he was forced to miss the next 14 games and played in four further games before missing the final 25 games due to a right knee sprain \citep{APgarnett,spears2009}. Thus considering data from one season in isolation does not give a complete picture of a player's ability.

\begin{table}
    \centering
    \begin{tabular}{l|c|c|c}\hline\hline
         & 2007--2009 data & 2009 data & 2007--2009 data \\
        Player Name & Model (\ref{eqn:newmodelH})--(\ref{eqn:newmodelA}) & Model (\ref{eqn:newmodelH})--(\ref{eqn:newmodelA}) & Model (\ref{sect:reducedvariance}) \\ \hline
        Steve Nash & 7.62 (1.53), 1 & 5.05 (1.77), 2 & 4.79 (1.68), 5 \\
        LeBron James & 7.08 (1.5), 2 & 4.96 (1.78), 3 & 6.47 (1.66), 1 \\
        Chris Paul & 6.87 (1.7), 3 & 4.51 (1.87), 5 & 5.02 (1.75), 4 \\
        Dwyane Wade & 6.69 (1.56), 4 & 5.28 (1.78), 1 & 5.59 (1.73), 2 \\
        Kobe Bryant & 5.91 (1.55), 5 & 3.51 (1.8), 11 & 4.15 (1.66), 8 \\
        Carmelo Anthony & 5.33 (1.52), 6 & 3.95 (1.75), 7 & 4.09 (1.72), 9 \\
        Dirk Nowitzki & 5.1 (1.52), 7 & 2.73 (1.77), 31 & 3.53 (1.64), 12 \\
        Pau Gasol & 4.89 (1.36), 8 & 4.28 (1.72), 6 & 4.15 (1.56), 7 \\
        Kevin Martin & 4.67 (1.42), 9 & 3.61 (1.66), 10 & 3.87 (1.63), 10 \\
        Michael Redd & 4.34 (1.68), 10 & 3.16 (1.91), 17 & 3.23 (1.88), 19 \\
        Danny Granger & 4.14 (1.4), 14 & 4.58 (1.64), 4 & 5.05 (1.59), 3 \\
        Brandon Roy & 4.08 (1.51), 15 & 3.91 (1.73), 8 & 4.18 (1.67), 6 \\
        Lamar Odom & 2.9 (1.39), 33 & 3.72 (1.62), 9 & 3.49 (1.54), 13 \\
    \end{tabular}
   \caption{\label{tab:offensiveratings} Centred offensive ratings showing mean (standard error), rank under model.}
\end{table}

\begin{table}
    \centering
    \begin{tabular}{l|c|c|c}\hline\hline
         & 2007--2009 data & 2009 data & 2007--2009 data \\
        Player Name & Model (\ref{eqn:newmodelH})--(\ref{eqn:newmodelA}) & Model (\ref{eqn:newmodelH})--(\ref{eqn:newmodelA}) & Model       (\ref{sect:reducedvariance}) \\ \hline
        Kevin Garnett & 4.07 (1.31), 1 & 2.47 (1.52), 9 & 2.51 (1.47), 8 \\
        Bruce Bowen & 4.04 (1.28), 2 & 2.62 (1.45), 3 & 2.26 (1.43), 16 \\
        Kurt Thomas & 3.47 (1.25), 3 & 2.29 (1.49), 13 & 1.92 (1.48), 28 \\
        Lamar Odom & 3.38 (1.21), 4 & 3.41 (1.39), 1 & 3.2 (1.34), 1 \\
        Chuck Hayes & 3.15 (1.37), 5 & 0.93 (1.58), 78 & 0.69 (1.57), 127 \\
        LeBron James & 3.09 (1.29), 6 & 3.18 (1.48), 2 & 2.78 (1.41), 4 \\
        Nene Hilario & 2.99 (1.38), 7 & 1.89 (1.52), 27 & 2.51 (1.45), 7 \\
        Amir Johnson & 2.92 (1.58), 8 & 2.49 (1.6), 8 & 2.26 (1.59), 15 \\
        Anderson Varejao & 2.75 (1.27), 9 & 1.15 (1.44), 58 & 1.27 (1.39), 59 \\
        Ron Artest & 2.64 (1.18), 10 & 2.27 (1.41), 14 & 1.91 (1.36), 29 \\
        Ime Udoka & 2.29 (1.29), 16 & 2.5 (1.53), 7 & 2.24 (1.52), 18 \\
        Andrew Bogut & 2.12 (1.47), 22 & 2.44 (1.63), 10 & 2.84 (1.62), 2 \\
        Jeff Foster & 2.09 (1.25), 24 & 2.23 (1.44), 16 & 2.48 (1.42), 9 \\
        Kirk Hinrich & 2.08 (1.26), 25 & 2.58 (1.45), 5 & 2.23 (1.43), 19 \\
        Rashard Lewis & 1.81 (1.2), 36 & 1.99 (1.41), 23 & 2.53 (1.34), 6 \\
        Marko Jaric & 1.45 (1.36), 54 & 2.62 (1.66), 4 & 2.81 (1.66), 3 \\
        Ronald Murray & 1.25 (1.22), 68 & 1.95 (1.38), 26 & 2.42 (1.35), 10 \\
        Quinton Ross & 1.22 (1.31), 74 & 2.25 (1.52), 15 & 2.54 (1.51), 5 \\
        Jarvis Hayes & 1.17 (1.25), 78 & 2.56 (1.45), 6 & 2.18 (1.43), 21 \\
    \end{tabular}
   \caption{\label{tab:defensiveratings} Centred defensive ratings showing mean (standard error), rank under model.}
\end{table}

\begin{table}
    \centering
    \begin{tabular}{l|c|c|c}\hline\hline
         & 2007--2009 data & 2009 data & 2007--2009 data \\
        Player Name & Model (\ref{eqn:newmodelH})--(\ref{eqn:newmodelA}) & Model (\ref{eqn:newmodelH})--(\ref{eqn:newmodelA}) & Model (\ref{sect:reducedvariance}) \\ \hline
        LeBron James & 10.17 (1.98), 1 & 8.14 (2.32), 1 & 9.25 (2.18), 1 \\
        Dwyane Wade & 7.6 (2.05), 2 & 6.41 (2.31), 3 & 6.61 (2.25), 3 \\
        Kevin Garnett & 7.18 (2.02), 3 & 4.19 (2.36), 13 & 4.16 (2.28), 12 \\
        Chris Paul & 6.7 (2.23), 4 & 5.46 (2.42), 4 & 5.41 (2.28), 5 \\
        Steve Nash & 6.46 (2.03), 5 & 4.48 (2.31), 9 & 4.06 (2.21), 14 \\
        Kobe Bryant & 6.45 (2.05), 6 & 4.31 (2.34), 11 & 5.02 (2.16), 6 \\
        Lamar Odom & 6.28 (1.85), 7 & 7.14 (2.13), 2 & 6.69 (2.04), 2 \\
        Tim Duncan & 5.72 (2.08), 8 & 2.62 (2.3), 42 & 2.7 (2.2), 41 \\
        Dirk Nowitzki & 5.57 (2.01), 9 & 3.19 (2.31), 28 & 3.45 (2.15), 25 \\
        Rashard Lewis & 5.41 (1.85), 10 & 5.27 (2.18), 5 & 5.57 (2.06), 4 \\
        LaMarcus Aldridge & 4.78 (2.02), 11 & 3.39 (2.28), 25 & 4.32 (2.19), 10 \\
        Yao Ming & 4.07 (1.98), 21 & 4.4 (2.2), 10 & 3.64 (2.12), 22 \\
        Matt Bonner & 3.8 (2.04), 25 & 4.85 (2.24), 6 & 4.48 (2.18), 8 \\
        Ray Allen & 3.71 (1.94), 27 & 4.64 (2.3), 8 & 3.65 (2.13), 21 \\
        Danny Granger & 3.31 (1.87), 36 & 4.16 (2.16), 14 & 4.88 (2.11), 7 \\
        Nene Hilario & 3.14 (2.14), 40 & 4.78 (2.39), 7 & 3.97 (2.26), 16 \\
        Brandon Roy & 3.07 (2), 42 & 4.15 (2.27), 15 & 4.36 (2.19), 9 \\
    \end{tabular}
   \caption{\label{tab:combinedratings} Centred combined ratings showing mean (standard error), rank under model.}
\end{table}

In the Bayesian framework advocated in this article, it is of interest to compute the posterior probability that one player is stronger than another.
For players $A$ and $B$, the marginal joint posterior density of their respective combined abilities is multivariate Gaussian. It is therefore straightforward to compute the posterior probability that the combined ability of player $A$ is larger that that of player $B$. These probabilities were computed for a subset of players using the 2007--2009 data, the results are in Table \ref{table:postprob}. It is also straightforward to compute the posterior probability that a particular player is the best; this is most easily achieved by a Monte Carlo estimate, simulating directly from the Gaussian posterior. With 1000 draws from this density based on the the 2007--2009 data, LeBron James had the highest rating on 466 occasions, followed by Dwayne Wade (77) and Kevin Garnett (51), that is the respective posterior probabilities that these players were the best at the end of the 2009 season were 0.47, 0.08 and 0.05. Using only the 2009 data, the posterior probability that the top player was LeBron James was 0.21; Lamar Odom, 0.11 and Dwayne Wade, 0.06.

\begin{table}
    \centering
   \footnotesize
    \begin{tabular}{c|c|c|c|c|c|c|c|c|c|}
          & \begin{sideways}LeBron James\end{sideways} & \begin{sideways}Dwyane Wade\end{sideways} & \begin{sideways}Kevin Garnett\end{sideways} & \begin{sideways}Chris Paul\end{sideways} & \begin{sideways}Steve Nash\end{sideways} & \begin{sideways}Kobe Bryant\end{sideways} & \begin{sideways}Lamar Odom\end{sideways} & \begin{sideways}Tim Duncan\end{sideways} & \begin{sideways}Dirk Nowitzki\end{sideways} \\ \hline
        Dwyane Wade & 0.82 & $\cdot$ & $\cdot$ & $\cdot$ & $\cdot$ & $\cdot$ & $\cdot$ & $\cdot$ & $\cdot$ \\
        Kevin Garnett & 0.86 & 0.56 & $\cdot$ & $\cdot$ & $\cdot$ & $\cdot$ & $\cdot$ & $\cdot$ & $\cdot$ \\
        Chris Paul & 0.88 & 0.62 & 0.56 & $\cdot$ & $\cdot$ & $\cdot$ & $\cdot$ & $\cdot$ & $\cdot$ \\
        Steve Nash & 0.91 & 0.66 & 0.6 & 0.53 & $\cdot$ & $\cdot$ & $\cdot$ & $\cdot$ & $\cdot$ \\
        Kobe Bryant & 0.91 & 0.66 & 0.6 & 0.53 & 0.5 & $\cdot$ & $\cdot$ & $\cdot$ & $\cdot$ \\
        Lamar Odom & 0.92 & 0.68 & 0.63 & 0.56 & 0.53 & 0.52 & $\cdot$ & $\cdot$ & $\cdot$ \\
        Tim Duncan & 0.94 & 0.74 & 0.69 & 0.63 & 0.6 & 0.6 & 0.58 & $\cdot$ & $\cdot$\\
        Dirk Nowitzki & 0.95 & 0.76 & 0.71 & 0.65 & 0.62 & 0.62 & 0.6 & 0.52 & $\cdot$ \\
        Rashard Lewis & 0.96 & 0.79 & 0.74 & 0.67 & 0.65 & 0.65 & 0.63 & 0.55 & 0.52\\
    \end{tabular}
   \caption{\label{table:postprob} Posterior probability that player on top of table is stronger than players at the side of the table. Computed using the 2007--2009 data.}
\end{table}

\section{Player Statistics\label{sect:playerstats}}

As mentioned in the introduction, one common, but \emph{ad hoc}, means of rating NBA players is by using their individual game statistics. In this section it will be considered whether the 2009 game statistics can predict player ability and if so, which of the game--statistics are important.

Consider estimating the offensive ability of player $i$, $\alpha_i$. Given covariate information (the game statistics) for this player, $X_{ij}$, a simple linear model would take the form,
\begin{equation}\label{eqn:setuplm1}
   \alpha_i = \sum_ja_jX_{ij} + \sigma_a\epsilon_i,
\end{equation}
where $\epsilon_i\simiid\N(0,1)$ and $\sigma_a>0$ is the residual variance. One problem with this model is that the quantity $\alpha_i$ is unknown, however it can be estimated as described earlier using model (\ref{sect:reducedvariance}) for example. 

The approach advocated here is therefore to replace $\alpha_i$ in (\ref{eqn:setuplm1}) by the estimate, $\hat\alpha_i$, and to allow the variance of each of the observations to vary to account for the relative precision of the estimated $\hat\alpha_i$. This leads to a linear model of the form,
\begin{equation}\label{eqn:gamestatsmodel}
   \hat\alpha_i = \sum_ja_jX_{ij} + \mathrm{S.E.}(\hat\alpha_i)\nu_i,
\end{equation}
where $\nu_i\simiid\N(0,1)$ and $X_{ij}$ corresponds to the value of the explanatory variable $X_j$ in Table \ref{tab:WLSestimates} for player $i$. The quantities, $\hat\alpha_i$ and $\mathrm{S.E.}(\hat\alpha_i)^2$, can be read directly from the estimated vector of means and from the leading diagonal of the estimated covariance matrix. A similar model was also set up for estimating $\hat\beta_i$ as a linear combination of the covariates, but with coefficients $b_j$.

Under model (\ref{eqn:gamestatsmodel}), the variance of $\hat\alpha_i$ is $\mathrm{S.E.}(\hat\alpha_i)^2$. Since this is different for each player, the estimates of the parameters $a_j$ are given by weighted least squares. Covariates were chosen via backwards selection, starting from the saturated model, excluding independent variables clearly not linked to the dependent variable (for example mean points scored was excluded from the defensive model). An intercept term was included in the model, though the coefficient of this term is not reported as it does not affect rankings derived from the fitted values. 

Table \ref{tab:WLSestimates} presents the results from these two linear models as well as listing the explanatory variables. The potential covariates of `3 pointer\%', `blocks/40 mins' and `personal fouls/40 mins' were not found to be significant in either model. Standardising (centring and scaling) the outcome and explanatory variables makes it easier to identify which effects have the greatest influence on ability; the un--standardised coefficients are also provided in this table for ease of reference. For the offensive model, it is perhaps not surprising that the number of points scored and the number of assists are the most important predictors of ability. These results further suggest that not losing possession and rebounds are more important than free throw or field goal accuracy. 

\begin{table}
    \centering
    \begin{tabular}{l|c|c|c|c|}
         &  \multicolumn{2}{|c|}{Standardised} & \multicolumn{2}{|c|}{Not Standardised} \\
         Statistic & Off. (95\% C.I.) & Def. (95\% C.I.) & Off. & Def. \\ \hline
        $X_1$ field goal \% & 0.11 (0,0.22) & $\cdot$ & 2.63 & $\cdot$ \\
        $X_2$ free throw \% & 0.12 (0.02,0.22) & $\cdot$ & 1.62 & $\cdot$ \\
        $X_3$ turnovers/40 mins & -0.22 (-0.32,-0.12) & 0.12 (0.01,0.22) & -0.45 & 0.05  \\
        $X_4$ total rebounds/40 mins & 0.14 (0.03,0.25) & $\cdot$ & 0.08 & $\cdot$ \\
        $X_5$ assists/40 mins & 0.35 (0.25,0.46) & $\cdot$ & 0.28 & $\cdot$ \\
        $X_6$ points scored/40 mins & 0.49 (0.4,0.59) & $\cdot$ & 0.18 & $\cdot$ \\
        $X_7$ steals/40 mins & $\cdot$ & 0.15 (0.04,0.26) & $\cdot$ & 0.33  \\
    \end{tabular}
    \caption{\label{tab:WLSestimates} Results from model (\ref{eqn:gamestatsmodel}) derived from the $2009$ data. The table gives both standardised and raw coefficients.}
\end{table}

\begin{figure}[htbp]
 \begin{minipage}{0.48\textwidth}
      \includegraphics[width=0.9\textwidth,height=0.9\textwidth]{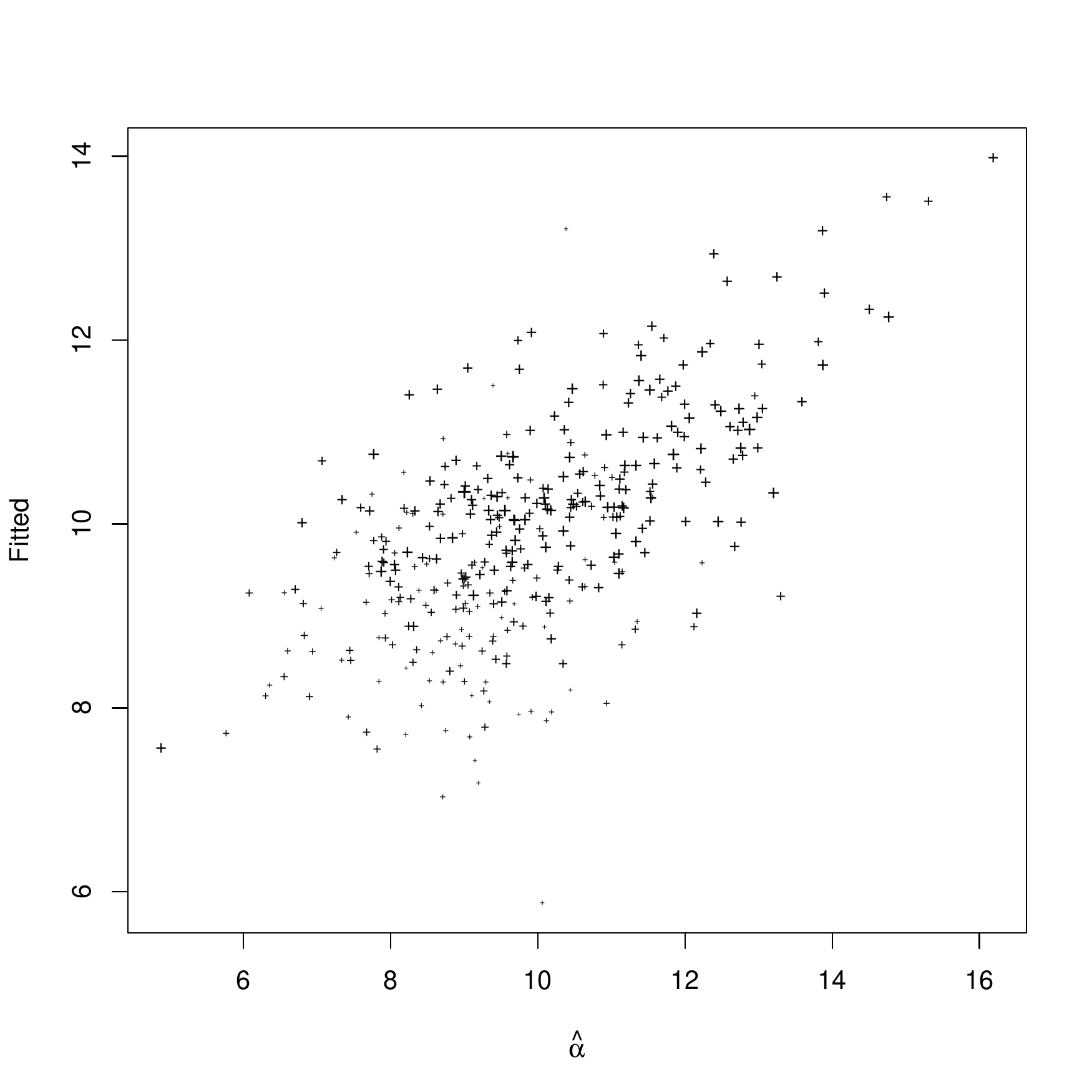}
   \end{minipage}
   \begin{minipage}{0.48\textwidth}
      \includegraphics[width=0.9\textwidth,height=0.9\textwidth]{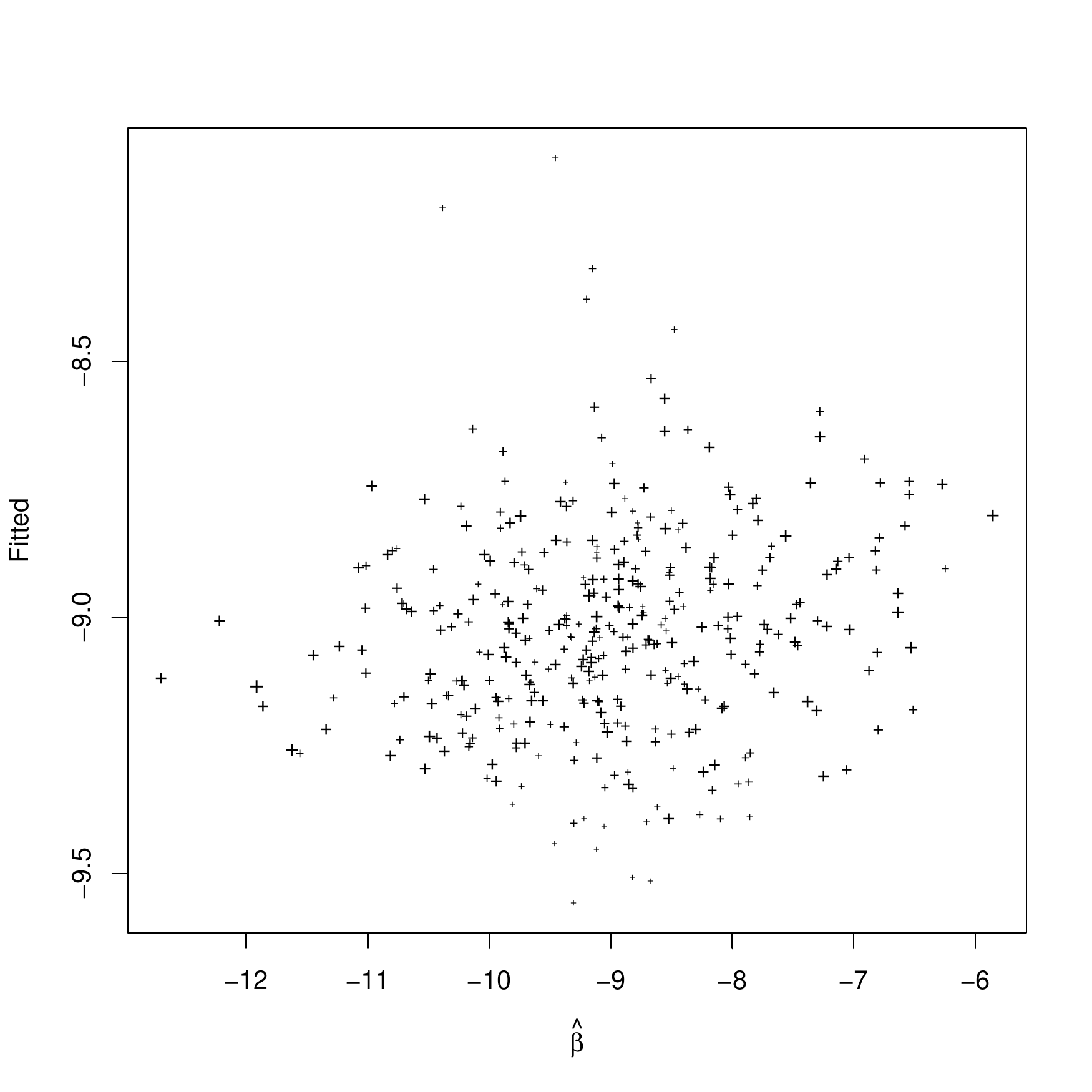}
   \end{minipage}
   \caption{\label{fig:gamestats} Plots of estimated ability against fitted values from the regression model (\ref{eqn:gamestatsmodel}). The markers are scaled according to the variance from (\ref{eqn:newmodelH})--(\ref{eqn:newmodelA}), with lager markers having less uncertainty from the original fit. The respective $R^2$ values for the offensive/defensive fits were $0.41$ and $0.03$.}
\end{figure}

Plots of the dependent variables against fitted values are shown in Figure \ref{fig:gamestats}, these plots show a clear evidence of a linear trend in the case of the offensive abilities, but not in the case of defensive ability. From the plot of the fitted offensive abilities, there is some evidence that the best players (in terms of the $\hat\alpha$s) also have much better game statistics than the rest -- consider the points in the top right corner of the plot. Of the ten game statistics considered, only steals, blocks and rebounds are potentially connected with defensive ability. There is no recorded information on, for example whether a fast--breaking team are contained by the forwards or guards. One possible problem with the fit of the defensive model may therefore be a lack of relevant covariate information.

Assuming that a linear model is appropriate for these data, the results in this table give the coefficients, $a_j$ and $b_j$, that can be used to predict player ability. For example in the case of the offensive model, the predicted strength of player $i$ is given by the expected value of $\hat\alpha_i$ under the model,
\begin{equation*}
   \E[\hat\alpha_i] = \sum_{j\in\{1,\ldots,6\}} a_jX_{ij},
\end{equation*} 
The fitted values from this linear model gives another means of rating and ranking players' offensive ability. The top ten offensive players using these fitted values are LeBron James, Chris Paul, Dwyane Wade, Gilbert Arenas, Kobe Bryant, Tony Parker, Dirk Nowitzki, Deron Williams, Brandon Roy, Steve Nash; seven of these players feature in Table \ref{tab:offensiveratings}.

\subsection{Offensive Ability of Forwards and Guards}

The three main types of position on court have different functionality in their offensive and defensive modes. For example, one aspect of the defensive ability of a guard is expertise in manoeuvres that prevent effective progress of the offense up the court -- they must me able to contend with players dribbling the ball. The function of the guards and forwards on the offense is also different: the former being more concerned with governing the general form of an attack and obtaining field goals, and the latter with shooting from the side of the court and obtaining rebounds \citep{ambler1979}. 

The linear model (\ref{eqn:gamestatsmodel}) was fitted to subsets of the players (all guards and all forwards separately) with the aim of finding any differences in covariate choice for predicting offensive ability. Using backward selection as above, the standardised covariate effects with standard errors in parenthesis are as follows. For the forwards, the most important ability were points scored, 0.47 (0.08); then assists, 0.44 (0.08); turnovers $-0.29$ (0.09); and lastly rebounds, 0.17 (0.08). For the guards the significant effects were points scored, 0.42 (0.07); personal fouls, $-0.23$ (0.07); then assists, $0.20$ (0.06); and lastly field goals, 0.19 (0.07). The $R^2$ values were respectively 0.40 and 0.47 for the forwards and guards. It is interesting to note that the different functions of the points and guards are reflected in the choice of covariates in these models: for the forwards, the ability to rebound is identified as important, whereas for the guards, shooting accuracy is relevant.


%


%
%

\section{Discussion\label{sect:discussion}}

This article introduces a new method for estimating both the offensive and defensive ability of NBA players and a justifiable way of conjoining this information to derive a combined estimate of player utility. To the knowledge of the authors, the model presented here is unique in providing a structured means of updating player abilities between years. One of the most important findings here is that whilst using player game statistics and a simple linear model to infer offensive ability may be okay, the very poor fit of the defensive ratings model suggests that defensive ability depends on some trait not measured by the current range of player game statistics.

At the end of each season, the NBA presents players with awards including `Most Valuable Player' (MVP), `Defensive Player of the Year', `Rookie of the Year' and `Most Improved Player' (MIP); though there is no material prize for the winning player, the awards are highly prestigious. The NBA employs a panel of sportswriters and broadcasters to rank their top five (in the case of the MVP award) or top three players. The individual rankings are combined by weighting each with respect to the voted position eg 10 points for first place votes and respectively 7, 5, 3, 1 for 2nd to 5th place \citep{NBA} and summing these over voters. Although the sportswriters and broadcasters are undoubtedly experts in their field, the rhetoric accompanying the announcement of the awards repeatedly refers to the much published player game statistics \cite{NBA-MVP}. It is of interest to compare the results presented here with those of the NBA committee.

The 2009 MVP according to model (\ref{sect:reducedvariance}) was LeBron James (Cleveland Cavaliers) since he has the highest estimated combined ability from Table \ref{tab:offensiveratings}, this is in agreement with the NBA decision. Similarly the best defensive player in 2009 was Lamar Odom (Los Angeles Lakers), here the NBA presented the award to Dwight Howard, who was ranked 34th by the new method. In order to rank rookie players, the approach advocated here is to use the combined ratings ratings from the 2007--2009 data, since in this setting the ability of non--rookie players is more accurately estimated by including historical data. The best rookie player here goes to Russell Westbrook of Oklahoma City Thunder (the award was given to Derrick Rose of the Chicago Bulls, ranked 27th by the new model). To compute the most improved player is a simple case of comparing the combined estimates of ability obtained from successive historical rankings, for example by using all available data up to 2008 and then all available data up to to 2009 -- excluding rookies, the largest difference between the combined estimates thus obtained gives the most improved player. Using the new method the 2009 MIP was Hakim Warrick of the Memphis Grizzlies (the NBA presented this award to Danny Granger of the Indiana Pacers, who was ranked 6th by the method described here).

\begin{singlespace}
  \bibliographystyle{Chicago}
  \bibliography{bibliography}

\begin{thebibliography}{}

\bibitem[\protect\citeauthoryear{Ambler}{Ambler}{1979}]{ambler1979}
Ambler, V. (1979).
\newblock {\em Basketball: The Basics for Coach and Player}.
\newblock Faber and Faber.

\bibitem[\protect\citeauthoryear{{Associated Press}}{{Associated
  Press}}{2009}]{APgarnett}
{Associated Press} (2009).
\newblock \url{http://espn.go.com/nba/recap?gameId=290219026}.

\bibitem[\protect\citeauthoryear{Berri}{Berri}{1999}]{berri1999}
Berri, D.~J. (1999).
\newblock Who is `most valuable'? measuring the player's production of wins in
  the {N}ational {B}asketball {A}ssociation.
\newblock {\em Managerial and Decision Economics\/}~{\em 20\/}(8), 411--427.

\bibitem[\protect\citeauthoryear{Entine and Small}{Entine and
  Small}{2008}]{entine2008}
Entine, O.~A. and D.~S. Small (2008).
\newblock The role of rest in the {NBA} home-court advantage.
\newblock {\em Journal of Quantitative Analysis in Sports\/}~{\em 4\/}(2).

\bibitem[\protect\citeauthoryear{{ESPN}}{{ESPN}}{2010}]{ESPNNBA}
{ESPN} (2010).
\newblock \url{http://espn.go.com/nba/}.

\bibitem[\protect\citeauthoryear{Harville}{Harville}{2003}]{harville2003}
Harville, D.~A. (2003, March).
\newblock The selection and seeding of college basketball or football teams for
  postseason competition.
\newblock {\em {Journal of the American Statistical Association}\/}~{\em
  98\/}(461), 17--27.

\bibitem[\protect\citeauthoryear{Ilardi and Barzilai}{Ilardi and
  Barzilai}{2008}]{ilardi2008}
Ilardi, S. and A.~Barzilai (2008).
\newblock \url{http://www.82games.com/ilardi2.thm}.

\bibitem[\protect\citeauthoryear{Kubatko, Oliver, Pelton, and
  Rosenbaum}{Kubatko et~al.}{2007}]{kubatko2007}
Kubatko, J., D.~Oliver, K.~Pelton, and D.~T. Rosenbaum (2007).
\newblock A starting point for analyzing basketball statistics.
\newblock {\em Journal of Quantitative Analysis in Sports\/}~{\em 3\/}(3), 1.

\bibitem[\protect\citeauthoryear{Macdonald}{Macdonald}{2010}]{macdonald2010}
Macdonald, B. (2010).
\newblock A regression--based plus--minus statistic for nhl players.
\newblock \url{http://arxiv.org/abs/1006.4310}.

\bibitem[\protect\citeauthoryear{{NBA}}{{NBA}}{2009}]{NBA-MVP}
{NBA} (2009).
\newblock
  \url{http://www.nba.com/2009/news/05/04/mvp.release.20090504/index.html}.

\bibitem[\protect\citeauthoryear{{NBA}}{{NBA}}{2010}]{NBA}
{NBA} (2010).
\newblock \url{http://www.nba.com/}.

\bibitem[\protect\citeauthoryear{Rosenbaum}{Rosenbaum}{2004}]{rosenbaum2004}
Rosenbaum, D.~T. (2004).
\newblock Measuring how {NBA} players help their teams win. {S}ource:
  \url{http://www.82games.com/comm30.htm}.

\bibitem[\protect\citeauthoryear{Spears}{Spears}{2009}]{spears2009}
Spears, M. (2009).
\newblock
  \url{http://www.boston.com/sports/basketball/celtics/articles/2009/05/18/com%
fortable_with_the_long_term_picture/}.

\bibitem[\protect\citeauthoryear{Zak, Huang, and Siegfried}{Zak
  et~al.}{1979}]{zak1979}
Zak, T.~A., C.~J. Huang, and J.~J. Siegfried (1979).
\newblock Production efficiency: The case of professional basketball.
\newblock {\em The Journal of Business\/}~{\em 52\/}(3), 379--392.

\end{thebibliography}
\end{singlespace}

\end{document}